\documentclass[traditabstract]{aa} % for the abstract without structuration 
\usepackage{graphicx}
\usepackage{amsmath}
\usepackage{natbib}
\usepackage{epsfig}
\usepackage{txfonts}
\newcommand{\mev}{\, \text{MeV}}
\newcommand{\gcc}{\, \text{g}/\text{cm}^{3}}

\begin{document}

\title{Electron fraction constraints based on Nuclear Statistical
  Equilibrium with beta equilibrium}
%\subtitle{}

\author{A.~Arcones \inst{1,2} \and G.~Mart\'inez-Pinedo \inst{2} \and
  L.~F.~Roberts \inst{3} \and S.~E.~Woosley \inst{3}}

\institute{Institut f\"ur Kernphysik, Technische Universit\"at
  Darmstadt, Schlossgartenstr.~9, D-64289 Darmstadt, Germany  \and GSI
  Helmholtzzentrum f\"ur Schwerionenforschung GmbH, Planckstr.~1,
  D-64291 Darmstadt, Germany  \and Department of Astronomy and
  Astrophysics, University of California, Santa Cruz, CA 95064}

\offprints{A. Arcones} 
\mail{a.arcones@gsi.de}

\date{received; accepted}      

%\date{\today}

\abstract{The electron-to-nucleon ratio or electron fraction is a key
  parameter in many astrophysical studies.  Its value is determined by
  weak-interaction rates that are based on theoretical calculations
  subject to several nuclear physics uncertainties. Consequently, it
  is important to have a model independent way of constraining the
  electron fraction value in different astrophysical
  environments. Here we show that nuclear statistical equilibrium
  combined with beta equilibrium can provide such a constraint. We
  test the validity of this approximation in presupernova models and
  give lower limits for the electron fraction in type Ia supernova and
  accretion-induced collapse.}


\titlerunning{Electron fraction constraints based on NSE and beta equilibrium} 
\maketitle

\section{Introduction}
\label{sec:introduction}

Weak interactions are known to play a critical role in the structure
and evolution of stars and the kind of nucleosynthesis they produce
\cite[e.g.][]{Langanke.Martinez-Pinedo:2003}.  For nuclei near the
valley of beta-stability, the relevant lifetimes of unstable nuclei
are well determined in the laboratory. However at the high
temperatures and densities found in stars and stellar explosions,
nuclei may be produced that are either very proton- or
neutron-rich. Usually the rates for weak interactions involving such
nuclei are only available from theory. Further, these nuclei exist in
a distribution of excited levels, making even a theoretical
calculation of their weak lifetimes problematic.

Over the years, different groups have calculated weak-interaction
rates for astrophysical applications \cite[for a review
see][]{Langanke.Martinez-Pinedo:2003}. In the 1980s Fuller, Fowler, \&
Newman \cite[][hereafter FFN]{Fuller.Fowler.Newman:1980,
  Fuller.Fowler.Newman:1982a, Fuller.Fowler.Newman:1982b,
  Fuller.Fowler.Newman:1985} calculated rates for electron capture,
positron capture, beta decay, and positron emission for
astrophysically relevant nuclei. Their tabulations were based on an
examination of all available information at that time and became the
standard in the field. With the improvement of methods and computers,
complete shell-model calculations were performed for $sd$-shell nuclei
\cite[]{Oda.Hino.ea:1994} and for $pf$-shell nuclei \cite[][hereafter
LMP]{Langanke.Martinez-Pinedo:2000,
  Langanke.Martinez-Pinedo:2001}. Although these new shell-model rates
and the FFN rates agree rather well for $sd$-shell nuclei, there are
appreciable difference at higher mass. The electron-capture rates, and
to a lesser extent, the beta-decay rates
\cite[see][]{Martinez-Pinedo.Langanke.Dean:2000} are smaller than the
FFN rates by approximately one order of magnitude.

In this paper we discuss an interesting limiting case that can be used
to test existing rate tabulations and their implementation as well as
to calculate astrophysical models in regions where rate information
may currently be inadequate.  {\sl Dynamic beta-equilibrium}
\cite[]{Tsuruta.Cameron:1965,Imshennik.etal:1967,Imshennik.Chechetkin:1971,
  Cameron:2001, Odrzywolek:2009} occurs when sufficient time elapses
at a given temperature and density for the electron mole number,
$Y_e$, to approach a constant. That is, for an ensemble of nuclei (but
not necessarily individual nuclei), electron capture and positron
emission occur at a rate balanced by electron emission and positron
capture. While physically not the same situation as when neutrino
absorption balances neutrino emission (thermal weak equilibrium), the
solution to the abundance equations is the same in many of the
astrophysical environments of interest, as we will show. Where this
approximation holds, a set of nuclear abundances can be obtained, and
thus a steady state value for $Y_e$, that depends only on nuclear bulk
properties (like binding energy and partition function), the
temperature, and the density. Similar work was done by
\cite{Imshennik.Chechetkin:1971}, but our results are based on
improved nuclear data present a broader overview of the possible
astrophysical scenarios where this approximation becomes useful.

There are only a few places in nature where thermal weak equilibrium
actually occurs, but they are interesting. Obviously the Big Bang is
one, but there the abundances are simple, just neutrons and protons.
Dynamic beta-equilibrium exists in neutron stars\footnote{This
  equilibrium is commonly denoted as chemical equilibrium or
  beta-equilibrium in the neutron stars literature}
\cite[]{Baym.Pethick.ea:1971,Weber:1999} and may occur briefly in the
cores of massive stars after silicon burning
\cite[]{Aufderheide.Fushiki.ea:1994a, Aufderheide.Fushiki.ea:1994b,
  Heger.Langanke.ea:2001, Heger.Woosley.ea:2001}. It may also happen
in white dwarfs that ignite carbon runaways at very high density
\cite[]{Canal.Isern.ea:1990}. In the latter case, there is thought to
exist a critical ignition density above which the reduction in
electron density by electron capture more than offsets the rise in
thermal pressure due to the burning. Above this density whose
  exact value depends on composition, initial mass, and accretion rate
  \cite[]{Canal.Isern.ea:1990}, it is expected that the white dwarf
collapses to a neutron star rather than exploding as a
supernova. This scenario is still only theoretical since it
  was not yet observed.  Just short of this critical density, which
may be $\sim8 \times 10^9$ g cm$^{-3}$ \cite[]{Timmes.Woosley:1992,
  Woosley:1997}, the star does explode but ejects neutron-rich
isotopes (e.g., $^{48}$Ca) that are difficult to make elsewhere in
nature \cite[]{Woosley:1997}.  In addition, as the
  Chandrasekhar mass scales as $M_{\mathrm{Ch}}\propto Y_e^2$, if a
  flame reduces the $Y_e$ before the star begins to expand vigorously,
  it will collapse. The race between expansion and electron capture
  must be determined by hydrodynamical simulations with physical
  fidelity, preferably in 2D or 3D.  Until they are done, it is
  difficult to make quantitative statements, but it is expected that a
  10\% change in the terminal value of $Y_e$ would cause an
  approximately 10\% change in the terminal density and might affect
  whether e.g., the critical ignition density for collapse is $8$ or
  $9 \times 10^9 \mathrm{g/cm}^3$
  \cite[]{Timmes.Woosley:1992}. Instabilities could make the
  dependence non-linear. Thus it is important to know what the minimum
  achievable value is for $Y_e$ at a given temperature and density.

To proceed, we calculate the electron fraction and the composition
assuming thermal weak equilibrium for temperatures and densities that
are relevant to various astrophysical environments. Thermal weak
equilibrium will only exist at such high temperatures and densities
that nuclear statistical equilibrium (NSE) also prevails, and we
assume that to be the case. We show that for temperatures, $T\lesssim
10$~GK, thermal weak equilibrium reduces approximately to the dynamic
beta-equilibrium of~\cite{Tsuruta.Cameron:1965}. This allows to
obtain results that depend only upon binding energies and partition
functions, not individual weak interaction rates. At higher
temperatures, thermal neutrinos are absorbed with a rate comparable to
the electron emission rate.  When this happens thermal weak
equilibrium and dynamic beta-equilibrium become
different. Nevertheless, the equilibrium value of the electron
fraction for dynamic beta-equilibrium can still be obtained using a
rather simple description of the relevant weak interaction rates.

Our paper is organized as follows. We start with a brief review of
NSE, thermal weak equilibrium and dynamic beta-equilibrium in
Sect.~\ref{sec:nse}. The results obtained from combining these
assumptions are presented in Sect.~\ref{sec:results} for broad ranges
of temperature and density. In Sect.~\ref{sec:applications}, we
discuss the astrophysical implications of our results for presupernova
models (Sect.~\ref{sec:presn}), proto-neutron star evolution
(Sect.~\ref{sec:proto-ns}), and accretion-induced collapse and type Ia
supernovae (Sect.~\ref{sec:typeIa}). Finally, we conclude in
Sect.~\ref{sec:conclusions}.

\section{NSE, thermal weak equilibrium and dynamic beta-equilibrium}
\label{sec:nse}
At high temperatures ($T > 4$~GK~$\approx 0.3$~MeV) the photon energy
in an astrophysical plasma is high enough to dissociate nuclei into
nucleons. At the same time, for densities $\rho \approx 10^{6}-10^{13}
\text{g/cm}^3$, rapid nuclear reactions lead to the formation of
nuclei. An equilibrium is reached between strong and electromagnetic
reactions. This is known as the NSE, where the abundances of nuclei
follow simply from Saha equation for a given density, temperature, and
electron fraction. Although this is well known, we briefly summarize
the relevant equations.

When NSE is reached, production and destruction of nuclei occur at the
same rate and the chemical potentials of nuclei and nucleons have to
satisfy:
\begin{equation}
 \mu_{(Z,A)} = Z\, \mu_p + (A-Z)\, \mu_n \, ,
 \label{eq:chem_equ}
\end{equation}
where the chemical potentials include the total mass
\begin{equation}
 \mu_i = m_i c^2 + kT \eta_i + \mu_c \, .
 \label{eq:chempot}
\end{equation}
Here $\eta_i$ is the degeneracy parameter, $k$ the Boltzmann constant,
$T$ the temperature, and $\mu_c$ the Coulomb contribution to the
chemical potential \cite[see
appendix in][]{Juodagalvis.Langanke.ea:2009}. The electron chemical
potential is given by $\mu_e = m_e c^2 +\eta_e$.

The abundance $Y_i=n_i/n$ of isotope $i$ is given by the ratio of its
number density $n_i$ over the total baryon number density $n$. The
number density $n_i$ is related to the degeneracy parameter $\eta_i$
assuming Maxwell-Boltzmann statistics by:
\begin{equation}
 n_i = \frac{G_i(T) e^{\eta_i}}{\Lambda_i^3} \, ,
 \label{eq:n_MB}
\end{equation}
with the partition function $G_i(T)=\sum(2J_k+1) e^{-E_k/kT}$ and the
thermal wavelength $\Lambda_i = \sqrt{2 \pi \hbar^2 /m_i kT}$.  For
the partition function we use values calculated by \cite{rauscher:2003}.

Protons also follow Maxwell-Boltzmann distributions over the
densities of interest, but neutrons become degenerate for $\rho
\gtrsim 10^{11} \,\text{g/cm}^3$ (the transition from the
non-degenerate to degenerate regime of course depends on the
temperature). Therefore, we use Fermi-Dirac distributions for nucleons
(both neutrons and protons), 
\begin{equation}
 n_i = \frac{8\pi \sqrt{2}}{h^3}m_i^3c^3 \beta_i^{3/2} [\mathcal{F}_{1/2}(\eta_i, \beta_i) 
 +\beta_i \mathcal{F}_{3/2}(\eta_i, \beta_i)] \, ,
 \label{eq:n_fermi}
\end{equation}
where $\beta=kT/mc^2$ is the so-called relativistic parameter, which
is almost zero for the range of temperatures considered here, and
$\mathcal{F}_k$ are Fermi functions:
\begin{equation}
 \mathcal{F}_k (\eta, \beta) = \int_0^{\infty} \frac{x^k (1+\frac{1}{2}\beta x)^{1/2} \text{d}x}
 {e^{-\eta+x}+1}\, . 
\end{equation}
The electron mole number is defined as $Y_e = n_e/n$, with the lepton
number density $n_e = n_{e^-} - n_{e^+}$.  The electron, $n_{e^-}$,
and positron, $n_{e^+}$, number densities are related to the
respective chemical potentials by Eq.~(\ref{eq:n_fermi}) with the
additional condition $\mu_e = \mu_{e^-} = - \mu_{e^+}$.

Combining Eqs.~(\ref{eq:chem_equ})~and~(\ref{eq:n_MB}), the
abundance of the nucleus with Z, A is
\begin{equation}
 Y(Z,A) = \frac{G_{(Z,A)}}{\rho/m_u}  \left( \frac{kT A m_u}{2 \pi \hbar^2}\right)^{3/2}  
 e^{(A-Z)\eta_n+Z\eta_p} e^{B/kT} e^{Z\mu_{c,p}-\mu_{c,(Z,A)}} \, .
 \label{eq:ynse}
\end{equation}
Here $\rho$ is the baryon density and $m_u$ the atomic mass unit. Note
that small corrections (less than $1\%$) arise from using $m_i=Am_u$
\cite[]{Seitenzahl.etal:2009}. To calculate the NSE composition, we
take the binding energies, $B=Zm_p c^2+(A-Z)m_nc^2-m_{(Z,A)}c^2$ from
\cite{Audi.Wapstra.Thibault:2003} and supplemented by theoretical
values from \cite{Moeller.Nix.ea:1995}.  In addition, the degeneracy
parameters $\eta_n$, $\eta_p$ are determined by charge neutrality and
mass conservation,
\begin{eqnarray}
 \sum Z_i Y_i  & = & Y_e \label{eq:charg_cons}\, ,\\
 \sum A_i Y_i   & = & 1.  \label{eq:mass_cons}
\end{eqnarray}
Eqs. (\ref{eq:ynse}) - (\ref{eq:mass_cons}) can be solved to
obtain $\eta_n$, $\eta_p$, and $Y(Z,A)$ for given temperature, density
and electron fraction.

The electron fraction is generally determined by weak process that are
much slower than the strong and electromagnetic nuclear reactions
responsible for maintaining NSE. However, if the system remains at a
given temperature and density long enough, weak reactions can also
reach equilibrium. If the neutrinos produced by the weak interaction
do not leave the system, the equilibrium reached is denoted as thermal
weak equilibrium. In order to determine the evolution of the system
before thermal weak equilibrium is achieved neutrino transport
calculations are necessary \cite[see][]{Murphy:1980}. Once, thermal
weak equilibrium is achieved, the neutrinos have a Fermi-Dirac
spectrum with a temperature, $T$, and neutrino chemical potential
$\mu_\nu = -\mu_{\bar{\nu}}$. The state of the system is completely
determined by the temperature, density and lepton fraction (relative
number of electrons and neutrinos).

Once thermal weak equilibrium is reached, the composition does not
change unless the temperature or density varies. Therefore, entropy is
conserved which implies that $ T \mathrm{d} S = \sum_i \mu_i
\mathrm{d}Y_i =0$ \cite[see][for useful discussion of equilibrium in
nucleosynthesis]{Meyer:1993}, thus
\begin{eqnarray}
 \mu_\nu \dot{Y}_\nu & + & \mu_e  \dot{Y}_e  +  \sum_i \mu_i
 \dot{Y}_i \nonumber\\ 
     & = &  (\mu_e -\mu_\nu )\dot{Y}_e + (\mu_p -\mu_n) \sum_i Z_i
     \dot{Y}_i + \mu_n \sum_i A_i \dot{Y}_i = 0 \, ,
\end{eqnarray}
here the right-hand side is obtained from the assumption of NSE and
the condition of lepton number conservation that implies, $\dot{Y}_\nu
= -\dot{Y}_e$.  The time variation of the abundance of nucleus $i$ is
given by $\dot{Y}_i$.  Using Eqs.~(\ref{eq:charg_cons})
and~(\ref{eq:mass_cons}) this reduces to
\begin{equation}
(\mu_e + \mu_p - \mu_\nu - \mu_n) \dot{Y}_e=0 . 
\label{eq:equilibrium}
\end{equation}
The value of $Y_e$ for which equilibrium is attained can be obtained
from the thermal weak equilibrium condition 
\begin{equation}
  \label{eq:theweeq}
  \mu_e + \mu_p = \mu_\nu + \mu_n.
\end{equation}
This implies that the value of the neutrino chemical potential is a
function of lepton fraction, temperature, and baryon
density. Alternatively, the equilibrium $Y_e$ can be obtained from
\begin{eqnarray}
\dot{Y}_e & = &  -\sum_i (\lambda^i_{\text{ec}} + \lambda^i_{\beta^+} +
\lambda^i_{\bar{\nu}_e} + \lambda^i_{\text{in}\beta^-}) Y_i  \nonumber\\
& & +\sum_i (\lambda^i_{\text{pc}} + \lambda^i_{\beta^-} + \lambda^i_{\nu_e} 
+ \lambda^i_{\text{in}\beta^+}) Y_i =0,
\label{eq:dotye}
\end{eqnarray}
which requires the knowledge of the weak interaction rates for
electron capture (ec), positron emission ($\beta^+$), antineutrino
absorption ($\bar{\nu}_e$), inverse of electron emission
(in$\beta^-$), positron capture (pc), electron emission ($\beta^-$),
neutrino absorption ($\nu_e$), and inverse of possitron emission
(in$\beta^+$) for a large ensemble of nuclei. Eqs.~(\ref{eq:theweeq})
and~(\ref{eq:dotye}) represent equivalent ways of obtaining the steady
state value of $Y_e$, if entropy is constant. Eq.~(\ref{eq:theweeq})
is certainly advantageous as it depends only on bulk nuclear
properties like binding energies and partition functions. However, its
applicability is rather limited because at stellar densities below
$\sim 10^{11}$~g~cm$^{-3}$ neutrinos scape and consequently a neutrino
chemical potential cannot be defined. In these cases, a dynamical
beta-equilibrium can still be achieved and its equilibrium $Y_e$ value
is given by Eq.~\eqref{eq:dotye} suppressing the neutrino absorption
rates:
\begin{equation}
 \label{eq:dotyedbe}
  \dot{Y}_e  =   -\sum_i (\lambda^i_{\text{ec}} + \lambda^i_{\beta^+} ) Y_i 
  +\sum_i (\lambda^i_{\text{pc}} + \lambda^i_{\beta^-} ) Y_i =0 .
\end{equation}
Nevertheless, we will show in the following that
Eq.~(\ref{eq:theweeq}) together with the assumption $\mu_\nu = 0$
provides a very good approximation to the solution of
Eq.~\eqref{eq:dotyedbe}, i.e., thermal weak equilibrium with $\mu_\nu
= 0$ reduces to dynamical beta-equilibrium. In this case the neutrino
density depends only on temperature:
\begin{eqnarray}
  \label{eq:nudens}
  n_\nu = n_{\bar{\nu}} & = & \frac{1}{2\pi^2\hbar^3}\int_0^\infty
  \frac{p^2}{1+\exp(pc/kT)} dp \nonumber \\
      & = & 7.61 \times 10^{27} \left(\frac{T}{\mathrm{GK}}\right)^3\
      \mathrm{cm}^{-3} \, ,
\end{eqnarray}
with $p$ being the neutrino momentum and $pc$ its energy.  At low
temperatures the neutrino densities are so small that can be neglected
and consequently the condition $\mu_\nu = 0$ reduces to $n_\nu =
0$. This is the situation in cold neutron stars where
Eq.~\eqref{eq:theweeq} with $\mu_\nu = 0$ is commonly used
\cite[]{Weber:1999} to determine the composition. However, at higher
temperatures the assumption of $\mu_\nu = 0$ results in neutrino
densities comparable to electron densities and consequently thermal
weak equilibrium, Eq.~\eqref{eq:theweeq}, and dynamical beta
equilibrium, Eq.~\eqref{eq:dotyedbe}, predict different values of
$Y_e$. Under these conditions, one has to solve
Eq.~\eqref{eq:dotyedbe} using a full set of weak interaction rates
\cite[]{Odrzywolek:2009}.

As discussed in the introduction, the determination of
weak-interaction rates on nuclei is a very challenging nuclear
structure problem that requires sophisticated many body
calculations. Reliable weak-interaction rates are necessary to
determine the evolution of the system and to account for the energy
loss by neutrinos \cite[]{Odrzywolek:2009}. However, in this paper we
are not interested in the evolution towards equilibrium, but only in
the value of the equilibrium electron fraction for a given temperature
and density. Therefore, only a simple description of the weak
interaction rates is necessary.

We start considering the case of thermal weak equilibrium under the
assumption that $\mu_{\nu}=0$. This has the advantage that we can use
Eqs.~(\ref{eq:theweeq}) and~(\ref{eq:dotye}) to check the validity of
our implementation of the weak interaction rates. Once these are
validated, we can can easily generalize our approach to the case with
neutrinos leaving the system, by setting the neutrino densities to
zero. Let us consider a couple of nuclei $(A,Z)$ and $(A,Z-1)$
connected by the weak interactions:

\begin{subequations}
 \label{eq:weakr}
 \begin{equation}
   \label{eq:weakec}
   e^- + (A,Z) \rightleftarrows (A,Z-1) + \nu_e,
 \end{equation}
 \begin{equation}
   \label{eq:weakpe}
   (A,Z) \rightleftarrows (A,Z-1) + e^+ + \nu_e, 
 \end{equation}  
 \begin{equation}
   \label{eq:weakee}
   (A,Z-1) \rightleftarrows  (A,Z)  + e^- + \bar{\nu}_e,
 \end{equation}
 \begin{equation}
   \label{eq:weakpc}
   e^+ + (A,Z-1) \rightleftarrows (A,Z) + \bar{\nu}_e. 
 \end{equation}
\end{subequations}
If neutrinos are present, these reactions can operate in both
directions.  Only the right direction is relevant for cases where
neutrinos leave the system.  One can approximate these reactions
assuming that they occur by a transition from the ground state of the
initial nucleus to the ground state of the final nucleus. This is an
exact result for neutrons and protons and has been shown to provide a
reasonable description of electron capture rates for a broad range of
nuclei~\cite[see figure 2
of][]{Langanke.Martinez-Pinedo.ea:2003}. Here we generalize this
prescription to the other rates in Eq.~(\ref{eq:weakr}). Our aim is
not to obtain accurate weak-interaction rates but an approach that
reproduces the balance between forward and reverse reactions once
equilibrium is achieved. In this sense our results should not be
considered as a substitute to the different tabulations of weak
interaction rates available in the literature. We use the following
expressions to calculate the weak-interaction rates in
Eq.~(\ref{eq:weakr}):

\begin{subequations}
 \label{eq:rates}
\begin{eqnarray}
 \lambda_{\text{ec}}&= C \int_{w_{\text{ec}}}^{\infty} E^2
 (E-Q)^2 &f_e(E,\mu_e)\nonumber \\ && (1-f_\nu(E-Q,\mu_\nu))dE, \label{eq:ec} \\
 \lambda_{\beta^+}&= C \int_{m_e c^2}^{-Q} E^2 (-Q-E)^2 &
 (1-f_p(E,-\mu_e))\nonumber \\ && (1-f_{\nu}(-Q-E,\mu_\nu)) d E, \label{eq:beta+} \\
 \lambda_{\beta^-}&= C \int_{m_e c^2}^{Q} E^2 (Q-E)^2 &
 (1-f_e(E,\mu_e)) \nonumber \\ && (1-f_{\bar{\nu}}(Q-E,-\mu_\nu)) d E, \label{eq:beta-} \\
 \lambda_{\text{pc}} &= C \int_{w_{\text{pc}}}^{\infty} E^2
   (E+Q)^2  &f_p(E,-\mu_e)\nonumber \\ && (1-f_{\bar{\nu}}(E+Q,-\mu_\nu)) dE, \label{eq:pc} \\
 \lambda_{\nu_e}&= C \int_{w_{\text{ec}}}^{\infty}
 E^2 (E-Q)^2 & (1-f_e(E,\mu_e)) \nonumber \\ &&f_{\nu}(E-Q,\mu_\nu) d E, \label{eq:nuabs}\\ 
 \lambda_{\text{in}\beta^+}&= C \int_{m_e c^2}^{-Q} E^2 (-Q-E)^2 &
 f_p(E,-\mu_e) f_{\nu}(-Q-E,\mu_\nu) d E, \label{eq:ibeta+}\\
 \lambda_{\text{in}\beta^-}&= C \int_{m_e c^2}^{Q} E^2 (Q-E)^2&
 f_e(E,\mu_e) f_{\bar{\nu}}(Q-E,-\mu_\nu) d E, \label{eq:ibeta-} \\
 \lambda_{\bar{\nu}_e}&= C \int_{w_{\text{pc}}}^{\infty}  E^2 (E+Q)^2 &
 (1-f_p(E,-\mu_e)) \nonumber \\ && f_{\bar{\nu}}(E+Q,-\mu_\nu) dE, \label{eq:nubarabs}
\end{eqnarray}
\end{subequations}
where the first four equations correspond to the forward direction in
Eq.~(\ref{eq:weakr}) describing the rates for electron capture (ec),
positron emission ($\beta^+$), electron emission ($\beta^-$), and
positron capture (pc), respectively. The last four equations describe
neutrino absorption ($\nu_e$), the inverse of positron emission
(in$\beta^+$), the inverse of electron emission (in$\beta^+$), and
antineutrino absorption ($\bar{\nu}_e$). $Q$ is the transition Q
value, i.e. $Q=M(A,Z-1)c^2-M(A,Z)c^2$, that is positive for
neutron-rich nuclei and protons. $M(A,Z)$ is the nuclear mass. The
quantity $C$ is given by

\begin{equation}
 \label{eq:a}
 C = \frac{B \ln 2 }{K (m_e c^2)^5},
\end{equation}
with $K=6144$~s~\citep{Hardy.Towner:2009} and $B$ the nuclear matrix
element. We use the value $B=1+3 g_A^2 = 5.76$ for nucleons and
$B=4.6$ for nuclei~\citep{Langanke.Martinez-Pinedo.ea:2003}. We have
checked that our results are not sensitive to 20\% variations of
this value. $f$ is the Fermi-Dirac distribution
\begin{equation}
 \label{eq:fd}
 f(E,\mu) = \frac{1}{1+\exp[(E-\mu)/kT]} \, .
\end{equation}
The quantity $w_{\text{ec}} = \max(m_e c^2, Q)$ is the threshold
energy for electron capture, and $w_{\text{pc}} = \max (m_e c^2, -Q)$
for positron capture rate. The integral in Eqs.~(\ref{eq:beta+})
and~(\ref{eq:ibeta+}) are defined only for $-Q> m_e c^2$. Similarly,
the integrals in Eqs.~(\ref{eq:beta-}) and~(\ref{eq:ibeta-}) are
defined only for $Q> m_e c^2$.

The value of the equilibrium electron fraction agrees up to 3
significant figures when we solve Eq.~(\ref{eq:theweeq}) or
Eq.~(\ref{eq:dotye}). This justifies our simple scheme,
Eq.~(\ref{eq:rates}), for calculating the weak interaction rates.

Dynamic beta-equilibrium, Eq.~(\ref{eq:dotyedbe}), can be obtained from
the rates in Eqs.~(\ref{eq:rates}) setting the neutrino distributions
to zero. The equilibrium electron fraction in dynamic beta-equilibrium
is typically smaller than the one calculated in thermal weak
equilibrium. As in the latter, neutrino absorption results in an
increase of the electron fraction.

\section{Results}
\label{sec:results}
\begin{figure}
 \centering
 \includegraphics[width=0.95\linewidth]{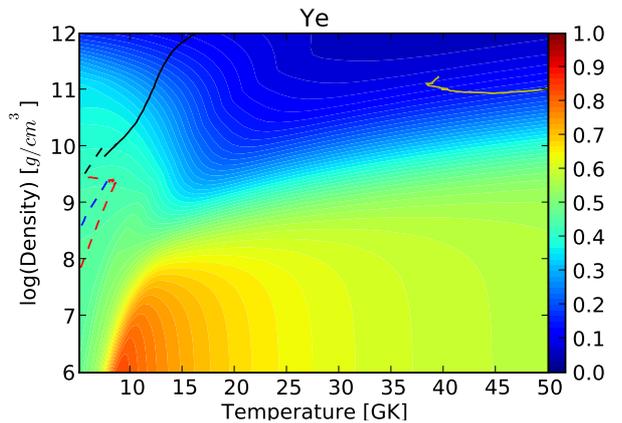}
 \caption{The electron fraction for NSE with thermal weak equilibrium
   is shown by color contours for a range of densities and
   temperatures where NSE can be applied and which are relevant in
   different astrophysical scenarios. The solid black line shows the
   core-collapse trajectory for a 15~$M_{\odot}$ (A.~Marek
   priv. com.); the dashed lines represent the presupernova evolution
   of a 15~$M_{\odot}$ (black) and 25~$M_{\odot}$ (blue) stars; the
   red dashed lines represent a type Ia trajectory (F.~R\"opke
   priv. com.); and the solid yellow lines correspond to the evolution
   of the proto-neutron star surface. Not all these scenarios will
   reach beta equilibrium and their electron fraction can therefore be
   different to the one shown here.}
 \label{fig:ye}
\end{figure}

We solve the NSE equations (Eqs.~(\ref{eq:ynse})-(\ref{eq:mass_cons}))
under the constraint of thermal weak equilibrium, Eq.~(\ref{eq:theweeq}),
assuming $\mu_\nu=0$ for temperatures from 5 to 150~GK ($0.43-13\mev$)
and densities in the range $10^{6}-10^{12} \gcc$. This results in the
equilibrium electron fraction shown in Fig.~\ref{fig:ye} and in the
composition shown in Figs.~\ref{fig:xa}, ~\ref{fig:nuclei}, and
~\ref{fig:lights}.  Because only small changes are expected at high
temperatures, we plot all quantities only over a smaller temperature
range.  The whole data table is available on request.

In Fig.~\ref{fig:ye} the color contours show the electron fraction
assuming thermal weak equilibrium and the lines represent trajectories
(density and temperature evolution) of various astrophysical
scenarios: presupernova, core collapse, proto-neutron star surface,
and type Ia supernovae.  The collapse trajectory (A.~Marek priv. com.)
of a 15~$M_{\odot}$ star based on the presupernova models of
\cite{Woosley02} is shown by the solid black line. When the inner core
of a massive star collapses the evolution is much faster than weak
reactions, therefore beta equilibrium is not achieved. Once densities
above $10^{11}$~g~cm$^{-3}$ are reached neutrinos begin to be trapped
and the assumption of $\mu_{\nu}=0$ is not valid any more. Therefore,
the equilibrium $Y_e$ differs from the value obtained in detailed
hydrodynamical simulations with neutrino transport. However, we
include this trajectory in our figure for completeness, to have an
idea of the range of temperatures and densities we are studying.
Presupernova trajectories for 15~$M_{\odot}$ and 25~$M_{\odot}$ stars
\cite[]{Woosley02} are shown by the dashed lines. This is discussed in
detail in Sect.~\ref{sec:presn}. The solid yellow line in the high
temperature region represents the evolution of the hot proto-neutron
star surface, where both thermal weak equilibrium and NSE are
fulfilled (Sect.~\ref{sec:proto-ns}). And the red dashed line
represents a trajectory of a type Ia supernova (F. R\"opke
priv. com.), where the evolution normally proceeds much faster than
the weak interaction timescale and consequently no equilibrium is
achieved (Sect.~\ref{sec:typeIa}).

\begin{figure}
 \centering
 \includegraphics[width=0.95\linewidth]{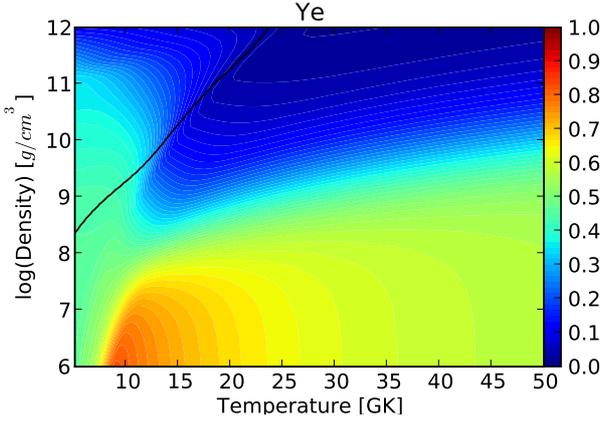}
 \caption{The electron fraction for NSE with dynamic beta-equilibrium
   is shown by color contours for the same range as in
   Fig.~\ref{fig:ye}. The black line represent the region where
   $\lambda_{\bar{\nu}_e} = \lambda_{\beta^-}$.}
 \label{fig:ye2}
\end{figure}

\begin{figure}
 \centering
 \includegraphics[width=0.95\linewidth]{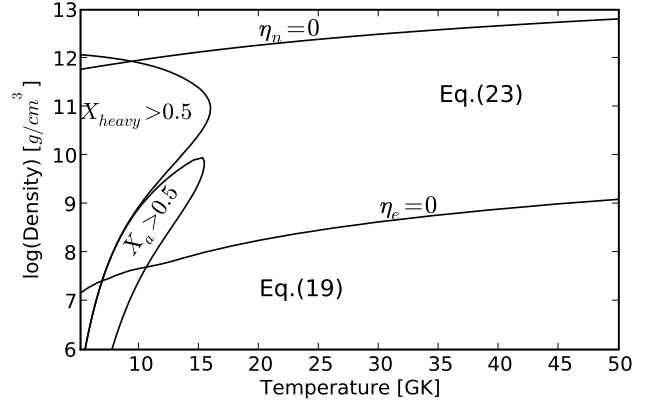}
 \caption{Lines limit the regions where different approximations for
   $Y_e$ can be used.  Horizontal lines indicate where the degeneracy
   of electrons ($\eta_e$) and neutrons ($\eta_n$) is zero. The
   regions where nuclei and alpha particles dominate the composition
   are at low temperatures.}
 \label{fig:ye_lines}
\end{figure}

\begin{figure}
   \includegraphics[width=0.95\linewidth]{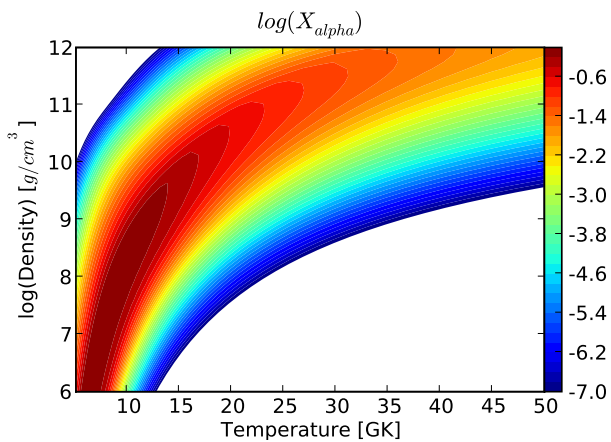}
   \caption{Mass fraction of alpha particles for the same temperature
     and density range as in Fig.~\ref{fig:ye}. In the white
       regions the mass fraction of alpha particles is negligible.}
 \label{fig:xa}
\end{figure}

\begin{figure*}
 \begin{tabular}{cc}
   \includegraphics[width=0.45\linewidth]{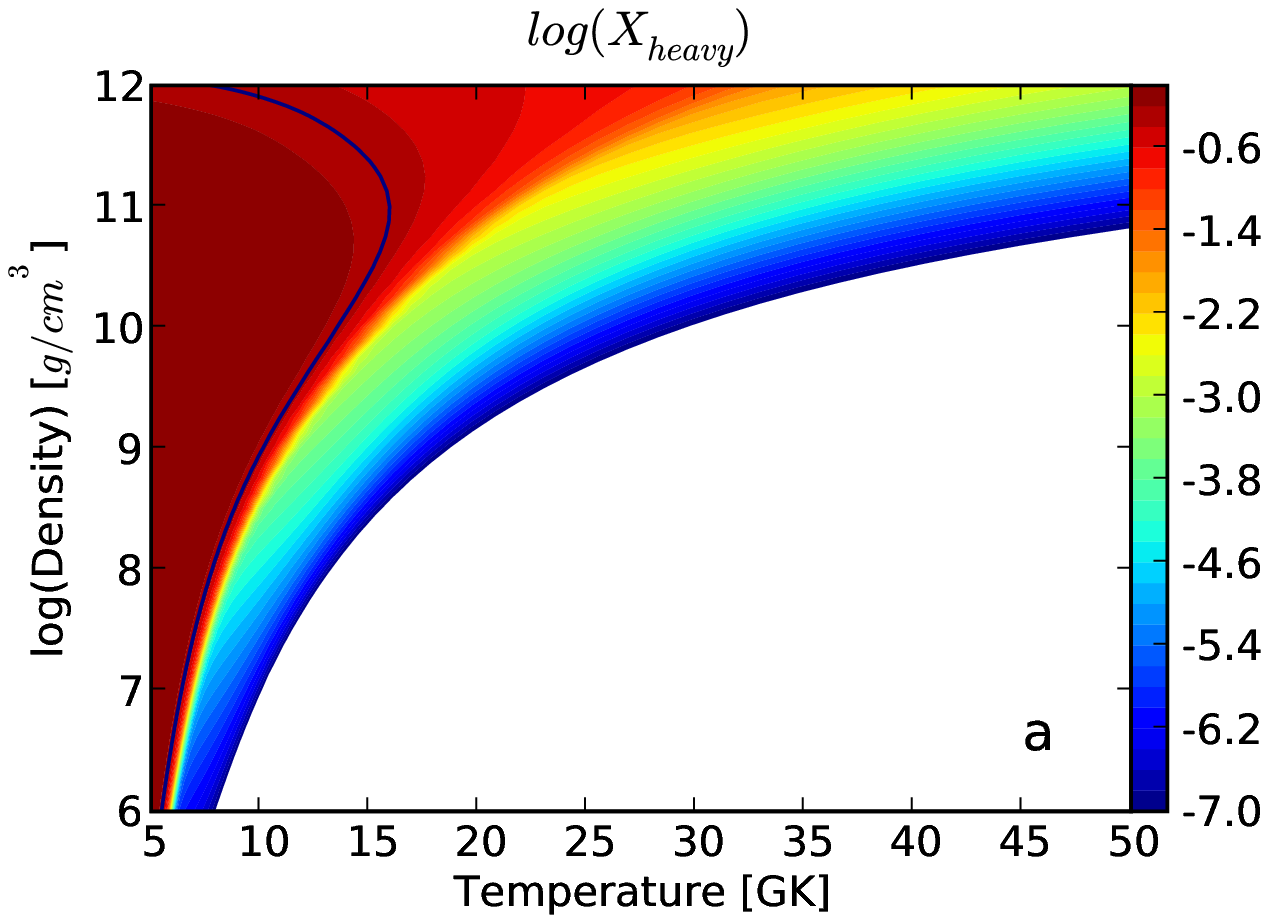}&
   \includegraphics[width=0.45\linewidth]{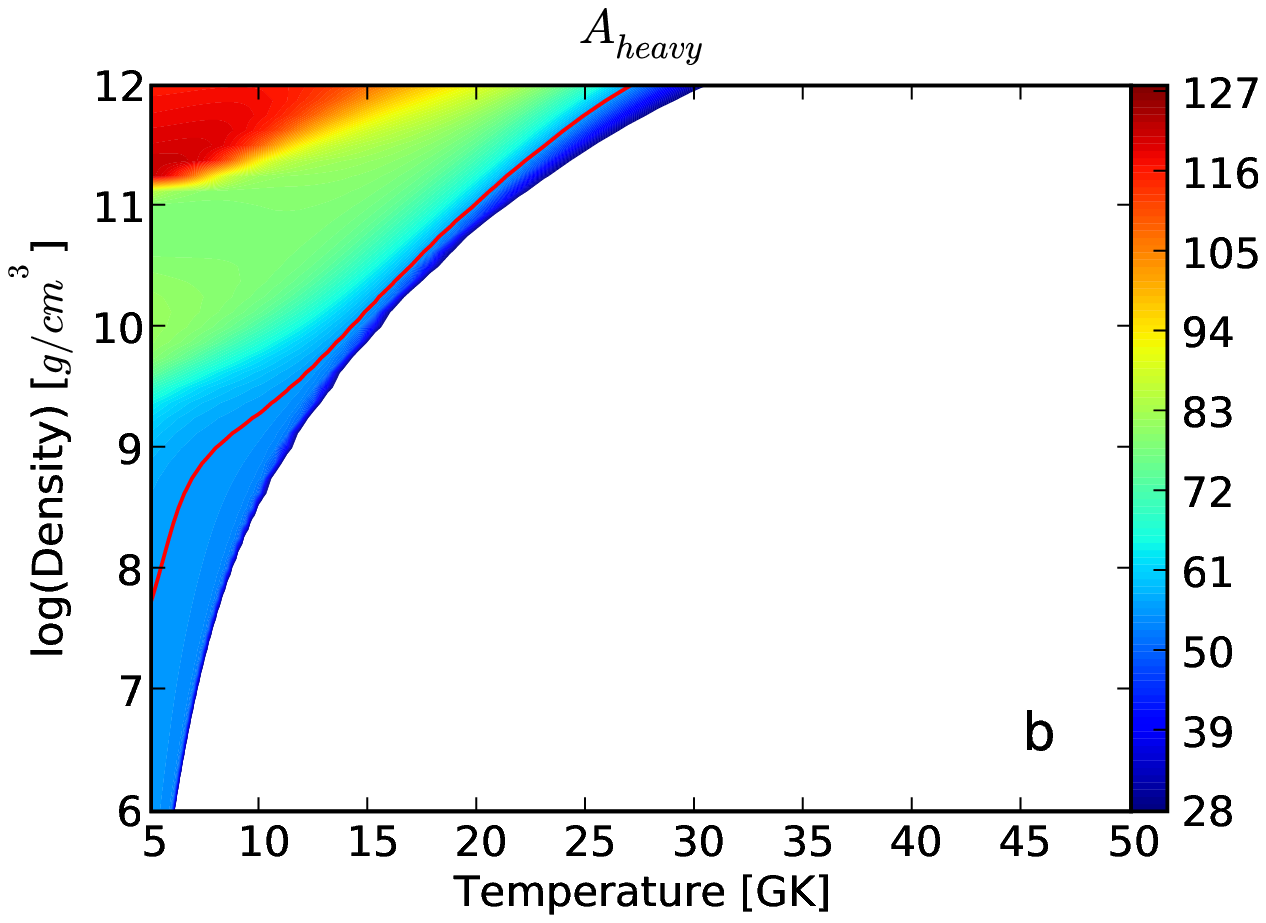}\\
   \includegraphics[width=0.45\linewidth]{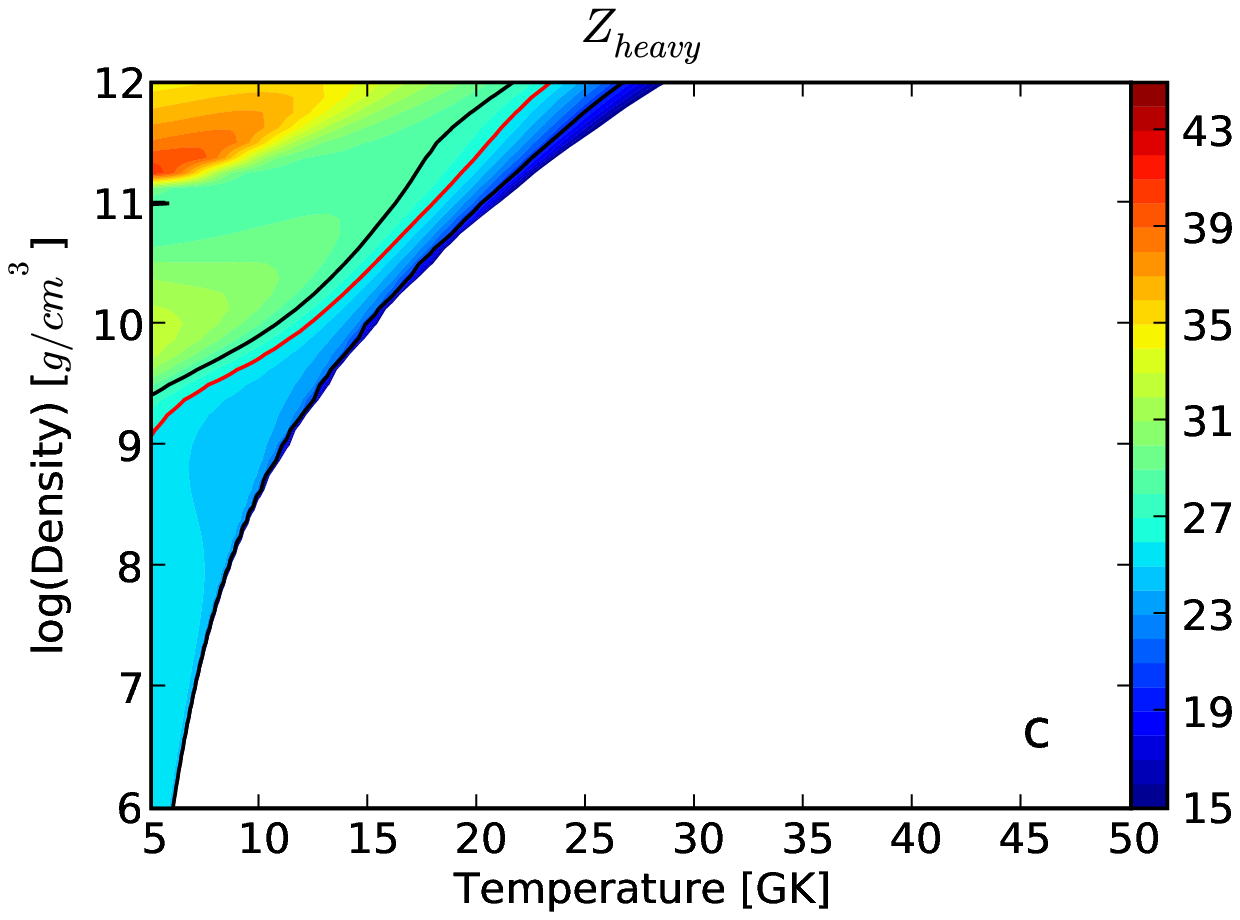}&
   \includegraphics[width=0.45\linewidth]{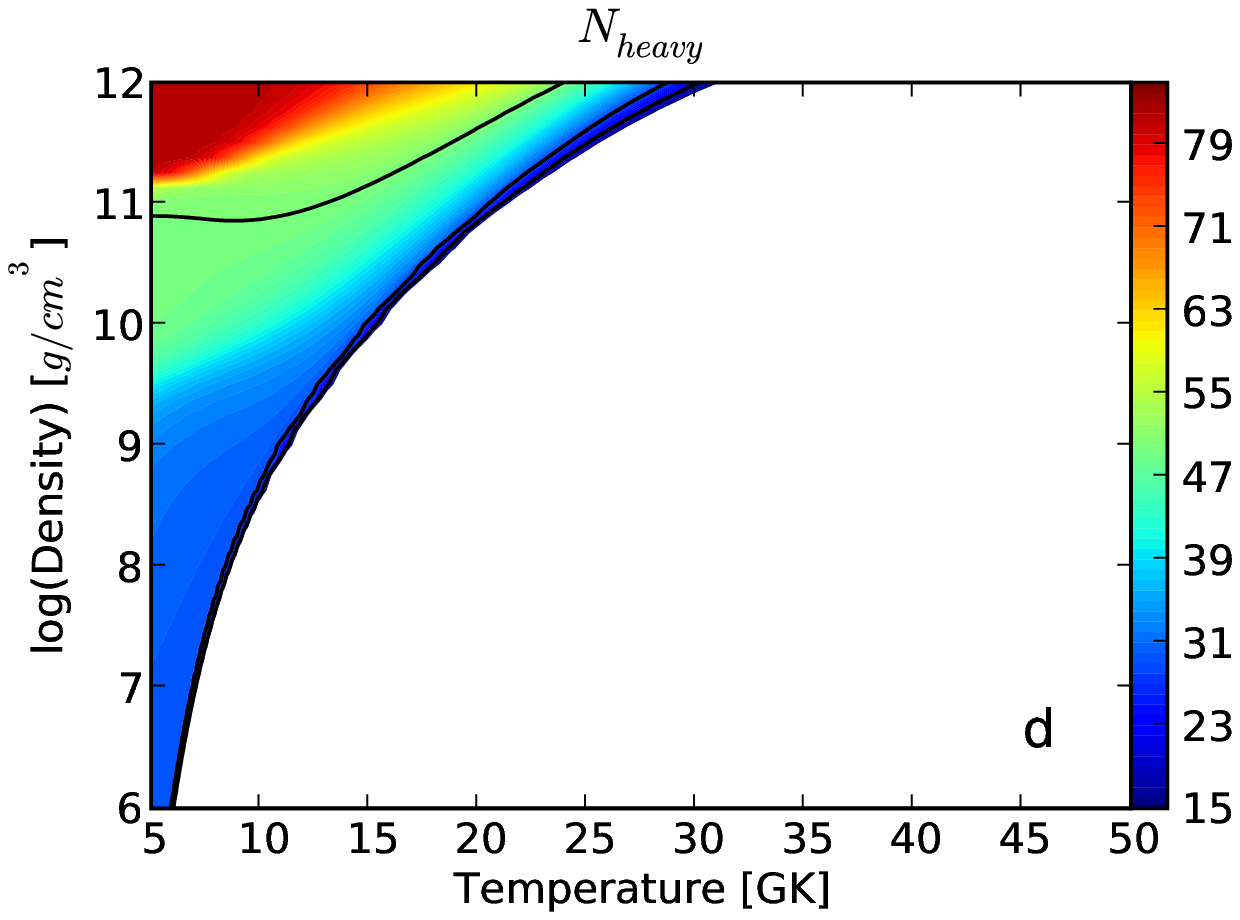}
 \end{tabular}
 \caption{The panel (a) shows the mass fraction of heavy nuclei
   (A~$>$~4), in the white region no heavy nuclei are present. The
   other panels give the average mass (b), atomic (c), and neutron (d)
   numbers. The blue line in the mass fraction plot marks the contour
   where heavy nuclei are half of the mass. In the average number
   plots, the black lines correspond to magic numbers and the red ones
   to $\,^{56}$Fe.}
 \label{fig:nuclei}
\end{figure*}

\begin{figure}
 \includegraphics[width=0.95\linewidth]{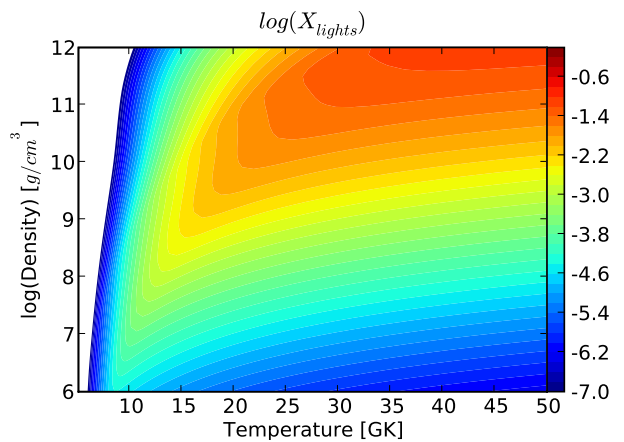}
 \caption{Mass fraction of light nuclei (A$<$4).}
 \label{fig:lights}
\end{figure}

For completeness we show in Fig.~\ref{fig:ye2} the electron fraction
obtained in NSE assuming dynamic beta-equilibrium,
Eq.~(\ref{eq:dotyedbe}), for the same temperatures and densities as
shown in Fig.~\ref{fig:ye}.  As discussed in Sect.~\ref{sec:nse}, we
expect that dynamic beta-equilibrium and thermal weak equilibrium with
$\mu_\nu=0$ give similar electron fraction whenever the absorption of
thermal neutrinos is negligible. This occurs on the left side of the
solid line shown in Fig.~\ref{fig:ye2}. This line marks the conditions
for which the antineutrino absorption would become equal to beta decay
if neutrinos did not escape.  At high temperatures the results for
both equilibria are not so different because the composition is
dominated by neutrons and protons and moreover positron captures also
become important.  In fact, at low densities (below the curve marked
as $\eta_e=0$ in Fig.~\ref{fig:ye_lines}) one has that $\mu_e \approx
\mu_{e^+}\approx 0$ implying that the contributions of
electrons/positrons and antineutrinos/neutrinos are similar. This
explains why both equilibria lead to same electron fraction for Big
Bang conditions.  The biggest differences between both approaches
appear in the region where the composition changes from heavy nuclei
to nucleons (Fig.~\ref{fig:ye_lines}). Moreover, it is in this region
where the electron fraction drops more abruptly and consequently its
value becomes rather sensitive to the equilibrium approach used.

Although solution of the full equations gives the value of the
equilibrium electron fraction, some simple approximations help to
understand the results. We divide the $\rho-T$ plane in different
temperature and density ranges to explain the behaviour of the
electron fraction. These ranges are shown in
Fig.~\ref{fig:ye_lines}. The two horizontal lines correspond to
electron and neutron degeneracy equal zero. We consider densities
below the line of $\eta_e=0$ to be low and above the line of
$\eta_n=0$ to be high. The temperatures are considered to be low when
heavy nuclei or alpha particles are present, which in the figure
correspond to the regions labeled as $X_{heavy}>0.5$ and $X_{a}>0.5$,
respectively.

In NSE at \emph{high temperature} ($T \gtrsim 15$~GK) one can assume
that only nucleons are present because all nuclei are dissociated,
thus $Y_e=Y_p$ and $Y_n=1-Y_e$. For \emph{low densities} conditions
are similar to the Big Bang, where the composition consists mainly of
non-degenerate neutrons, protons, and electrons. Therefore, we have
$\mu_e=0$ and $\mu_n \approx \mu_p$ that together with
Eq.~(\ref{eq:n_MB}) give:
\begin{equation}
 Y_e = \frac{1}{1+e^{-\Delta/kT}}\, ,
 \label{eq:ye_formula1}
\end{equation}
where $\Delta = m_n-m_p=1.293$~MeV. Under these conditions $Y_e$ can
reach rather high values when the temperature becomes smaller than
$\Delta$.  At higher densities ($\rho \gtrsim 10^8-10^9 \,
\mathrm{g/cm}^3$), electrons become degenerate and this approximation
is not valid anymore. This happens when the density is above:
\begin{equation}
 \rho \geq 1.132 \cdot 10^{7} \frac{(kT)^{3/2}}{Y_e} (0.6773 + 2.256 \, kT) \, ,
 \label{eq:dens_etae0}
\end{equation}
which corresponds to Eq.~(\ref{eq:n_fermi}) assuming $\eta_e \approx
0$ and using the values from \cite{Bludman.vanRiper:1977} for the
Fermi functions ($\mathcal{F}_{1/2}(\eta=0)=0.6773$ and
$\mathcal{F}_{3/2}(\eta=0)=1.153$). This line is shown in
Fig.~\ref{fig:ye_lines} labeled as $\eta_e=0$.

For densities above the line given by Eq.~(\ref{eq:dens_etae0}) the
electron fraction can be estimated by taking into account that the
electron number density is approximately given by:
\begin{equation}
 n_e  = 8\pi \left( \frac{kT}{hc} \right)^3 \frac{1}{3} \eta_e^3 \, .
 \label{eq:ne}
\end{equation}
From this equation we get directly $\eta_e$. If we assume that the
nucleons follow Maxwell-Boltzmann distributions (Eq.~(\ref{eq:n_MB})),
their chemical potentials are:
\begin{equation}
 \mu_{n(p)} = m_{n(p)} c^2 + kT \text{ln}\left( \frac{n_{n(p)} \Lambda_{n(p)}^3}{G} \right) \, .
\end{equation}
Inserting all this in the beta equilibrium equation ($\mu_n =
\mu_p+\mu_e$) we obtain a non-linear equation for the electron
fraction:
\begin{equation}
 \Delta - m_e c^2 +kT \text{ln}\left( \left(\frac{m_p}{m_n} \right)^{3/2}\frac{1-Y_e}{Y_e} \right)
- \left( \frac{3Y_e \rho}{8 \pi m_u} \right)^{1/3} \frac{hc}{kT} =0 \, .
 \label{eq:ye_formula2}
\end{equation}
This can be solved numerically using as guess value the $Y_e$ of
Eq.~(\ref{eq:ye_formula1}).  At high density, low $Y_e$ (more
neutron-rich material) is expected from
Eq.~(\ref{eq:ye_formula2}). The last term in
Eq.~(\ref{eq:ye_formula2}) increases as the density increases, unless
$Y_e$ decreases. Moreover, a reduction of $Y_e$ leads to an increase
of the third term, which has opposite sign to the term containing the
density. The exact solution, shown in Fig.~\ref{fig:ye}, demonstrates
the neutron richness of the matter at high densities.

For \emph{low temperatures} ($T\lesssim 15$~GK) the situation is more
complicated because of the presence of bound nuclei.  We estimate the
densities and temperatures at which nuclei dominate the composition
following the derivation of \cite{Hoyle.Fowler:1960}. This leads to
nuclei (represented by $\,^{56}$Fe) corresponding to half of the mass
at:
\begin{equation}
 \log  \rho = 11.62 + 1.5 \log T_9 - \frac{39.17}{T_9} .
 \label{eq:rho56Fe}
\end{equation}
In a similar way, taking mass fraction ($X_i=A_i Y_i$) of alpha
particles $X_{\alpha}=0.5$, alpha particles represent half of the
mass, for which one has:
\begin{equation} 
 \log \rho  =  29.68 + 4.5 \log T_9 + \log \frac{X_{\alpha}}{X_p^2X_n^2} -\frac{142.62}{T_9} \, ,
\label{eq:rhoalpha}
\end{equation}
where $X_{\alpha}$, $X_{n}$, and $X_{p}$ are the alpha, neutron, and
proton mass fractions, respectively, which are obtained from mass and
charge conservation:
\begin{eqnarray}
 1    & = & X_{\alpha} + X_p + X_n \, ,\\
 Y_e  & = & \frac{1}{2}X_{\alpha} + X_p \, .
\end{eqnarray}
At \emph{low densities}, between the two lines given by
Eqs.~(\ref{eq:rho56Fe}) and (\ref{eq:rhoalpha}), there are mainly
alpha particles, therefore $Y_e=0.5$. Above the line given by
Eq.~(\ref{eq:rho56Fe}), the composition is more complicated due to the
presence of different nuclei. For \emph{high density} the nuclei
become neutron rich. We can delimit the region where heavy nuclei are
present using simple approximations, as described before. However, the
detail composition and thus the electron fraction in this region can
only be obtained by numerically solving NSE equations under the
assumption of thermal weak equilibrium or dynamic beta-equilibrium, as
described in Sect.~\ref{sec:nse}.

The mass fraction of alpha-particles is shown in
Fig.~\ref{fig:xa}. Alpha-particles appear mostly at low density and
temperature, between the region where nucleons dominate the
composition (high temperature) and the region where heavier nuclei are
formed from the alpha particles. Notice that the formation alpha
particle leads to a change in the behaviour of $Y_e$ at low densities
(Fig.~\ref{fig:ye}). The mass fractions of heavy nuclei ($A>4$) are
shown in the panel (a) of Fig.~\ref{fig:nuclei}, they dominate the
composition at low temperatures and high
densities. Figure~\ref{fig:nuclei} shows the average mass, proton and
neutron numbers of the nuclei in panels (b), (c), and (d)
respectively.  In addition we find that the abundances of light nuclei
(A~$<4$) are significant at high temperatures and relative high
densities, as shown in Fig.~\ref{fig:lights}. Such conditions are
found in the outer layer of the proto-neutron star which is still hot
as shown by the trajectory (yellow line) in Fig.~\ref{fig:ye}.

\section{Discussion and astrophysical implications}
\label{sec:applications}
There are several astrophysical environments where our results are
interesting (Fig.~\ref{fig:ye}). These are situations where sufficient
time elapses at high temperature for NSE to prevail and for weak
interactions to approach equilibrium. Even when equilibrium is not
fully achieved, our results give an interesting lower bound on $Y_e$
in cases where electron capture is dominant.

\subsection{Presupernova evolution}
\label{sec:presn}
\begin{figure*}
 \begin{tabular}{cc}
   \includegraphics[width=0.45\linewidth]{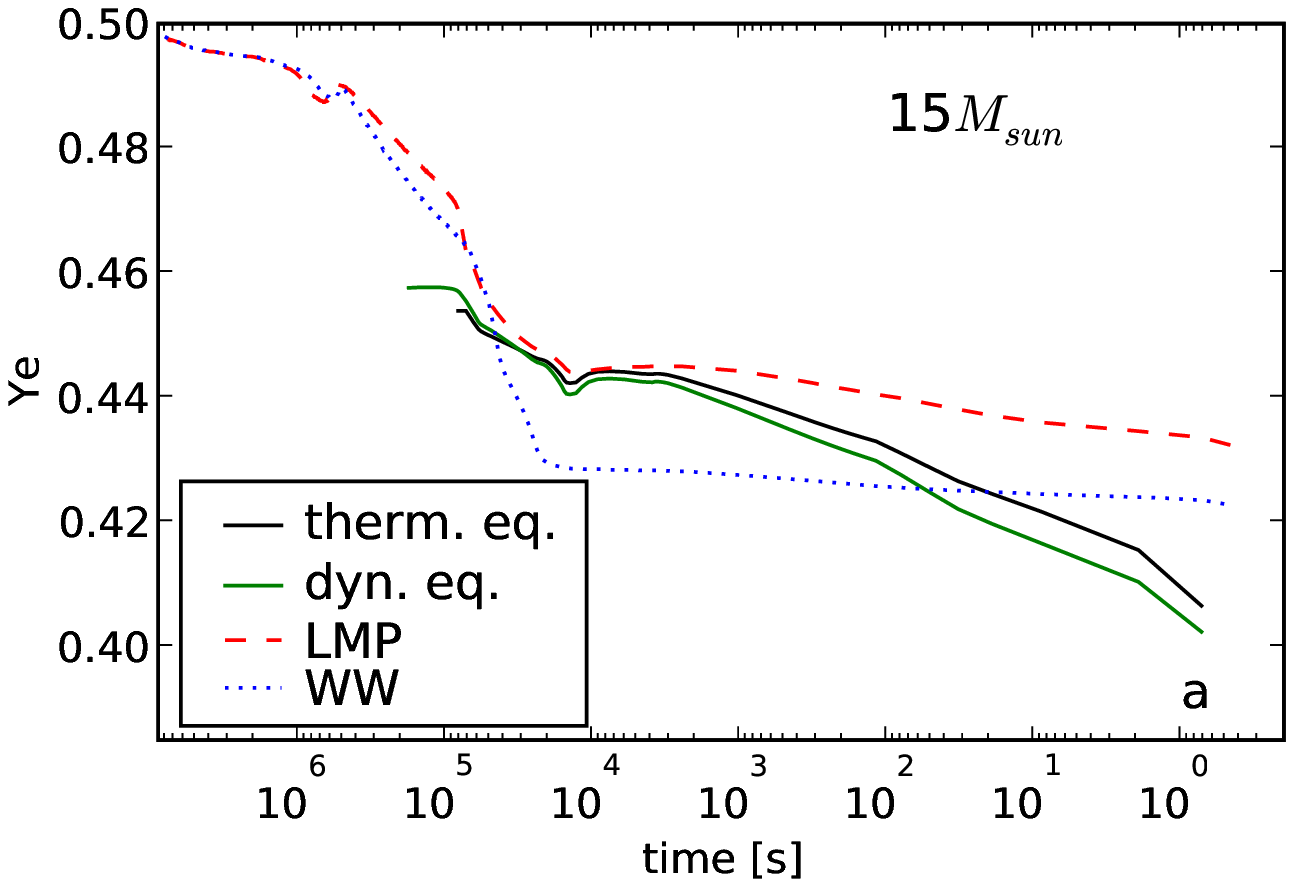}&
   \includegraphics[width=0.45\linewidth]{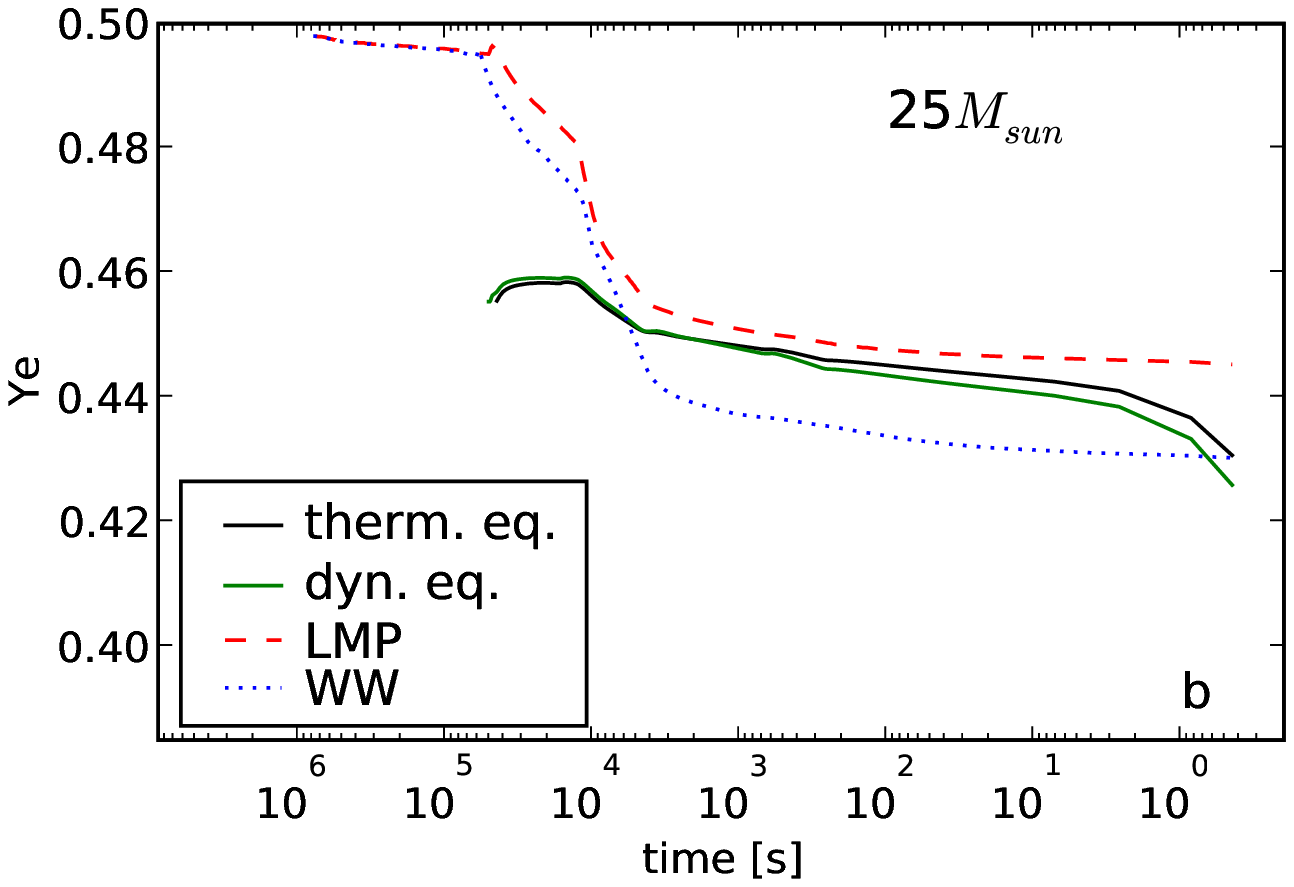}\\
   \includegraphics[width=0.45\linewidth]{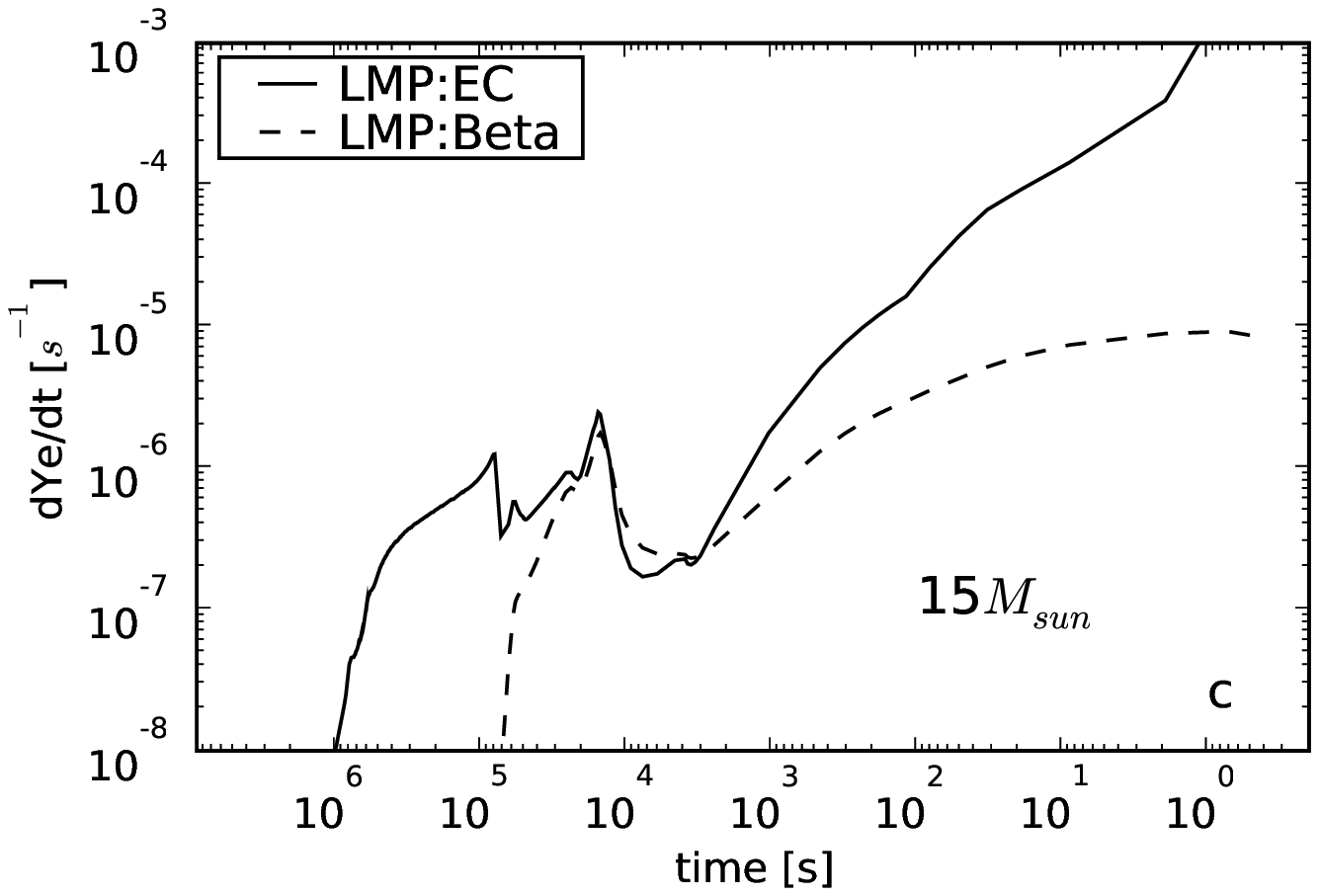}&
   \includegraphics[width=0.45\linewidth]{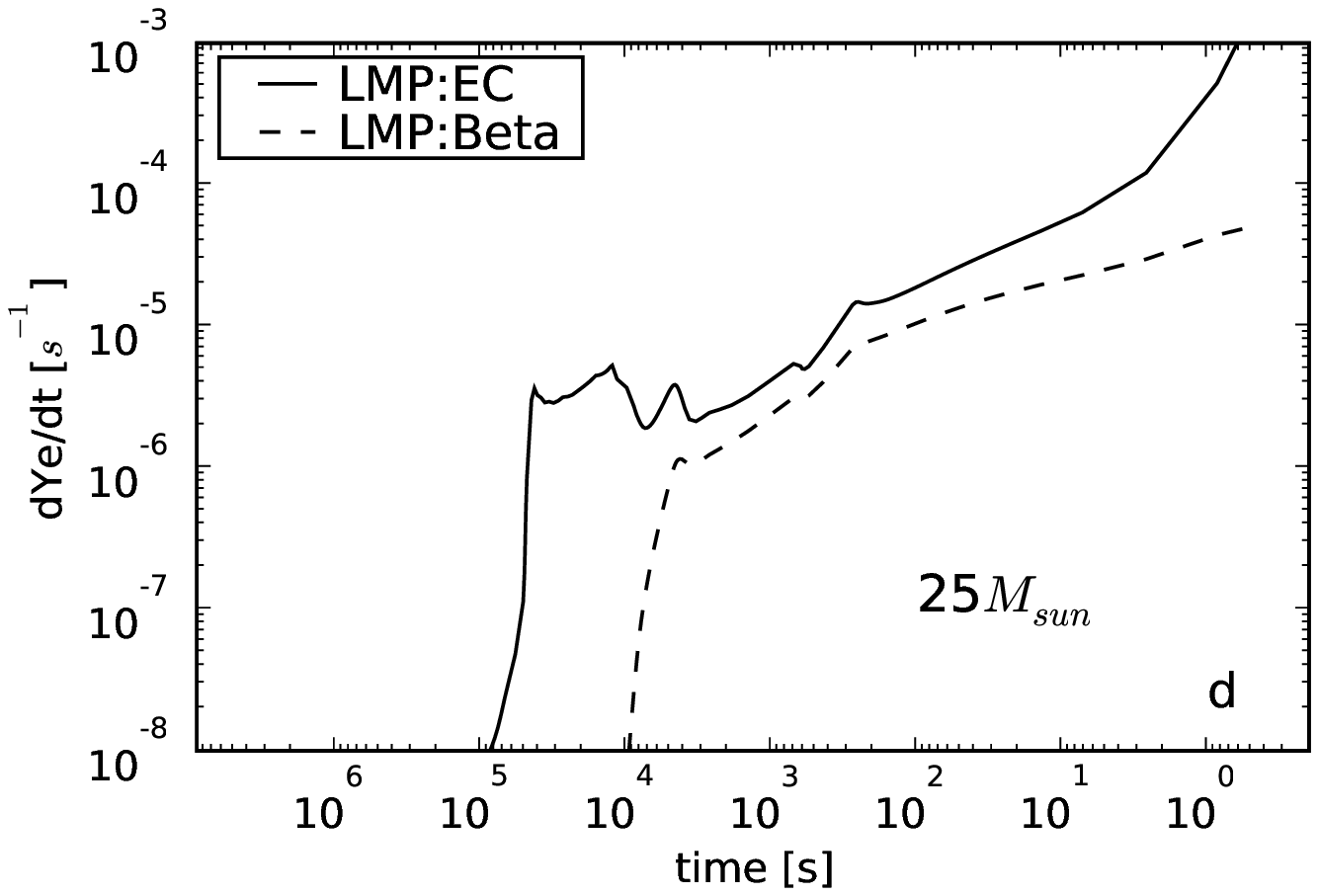}
 \end{tabular}
 \caption{The electron fraction (upper plots), the electron-capture,
   and beta-decay rates (bottom plots) as a function of time till
   collapse are shown for two presupernova models: 15~$M_{\odot}$
   (left) and 25~$M_{\odot}$ (right). In the electron fraction plots
   the dotted lines (blue) and dashed lines (red) were already shown
   by \cite{Heger.Langanke.ea:2001} and correspond to the old models
   of \cite{Woosley.Weaver:1995} and to the new ones of
   \cite{Heger.Woosley.ea:2001}, respectively. Here we have added the
   solid lines, which results from assuming NSE and thermal weak
   equilibrium (solid black line) or dynamic beta-equilibrium (solid
   green line).}
 \label{fig:presn}
\end{figure*}

We investigate here two presupernova models that were already studied
in detail in the frame of weak interactions by
\cite{Heger.Langanke.ea:2001,Heger.Woosley.ea:2001}. Fig.~\ref{fig:presn}
shows the results for the 15~$M_{\odot}$ star in the left column and
for the 25~$M_{\odot}$ star in the right one. All results are plotted
versus time before collapse which is at time zero.  The dotted lines
are the $Y_e$ of the old models of \cite{Woosley.Weaver:1995}
(WW). These models were computed with electron-capture rates of
\cite{Fuller.Fowler.Newman:1980} (FFN) and beta-decay rates of
\cite{Hansen:1968, Mazurek.Truran.Cameron:1974}. The dashed lines were
obtained when using the weak-interaction rates of
\cite{Langanke.Martinez-Pinedo:2001} (LMP) in the calculation of the
presupernova models. These two results were already presented in
\cite{Heger.Langanke.ea:2001,Heger.Woosley.ea:2001}, where they
discuss that differences in the electron fraction are mainly due to
discrepancies between the old beta decay rates and the new ones. The
bottom panels in Fig.~\ref{fig:presn} show the electron-capture and
beta-decay rates as calculated by \cite{Langanke.Martinez-Pinedo:2001}
and used for the presupernova models presented in
\cite{Heger.Langanke.ea:2001, Heger.Woosley.ea:2001}.

Before silicon burning, the evolution of the electron fraction is very
similar for WW and LMP because electron captures dominate over beta
decays (see bottom panels in Fig.~\ref{fig:presn}) and these rates are
similar for FFN and LMP. In this first phase when $Y_e$ starts to
decrease from the starting value of 0.5, the temperature is too low
and the assumption of NSE is not valid and there are no values for the
equilibrium electron fraction.

During silicon shell burning the electron fraction continues dropping
and beta decay becomes important and comparable to electron capture.
For the 15~$M_{\odot}$ star both rates are equal and a dynamic
beta-equilibrium is reached \cite[]{Aufderheide.Fushiki.ea:1994b,
  Heger.Woosley.ea:2001}. After silicon burning the electron fraction
stays almost constant although electron captures become dominant over
beta decays because the dynamical evolution is faster than the capture
rates. Therefore, as it was pointed out by
\cite{Heger.Woosley.ea:2001}, the final electron fraction is set
before the core begins its final contraction. The most important
period for determining the core structure is thus the silicon shell
burning phase.

We use the density and temperature evolutions shown in
Fig.~\ref{fig:ye} for the presupernova models of
\cite{Woosley.Weaver:1995} and calculate the equilibrium electron
fraction. This $Y_e$ correspond to the solid lines in the upper panels
of Fig.~\ref{fig:presn}. Notice that there are two solid lines, the
black one results from the assumption of thermal weak equilibrium,
Eq.~(\ref{eq:theweeq}), while the green one is obtained under the
assumption of dynamic beta-equilibrium, Eq.~(\ref{eq:dotyedbe}). The
resulting $Y_e$ values are almost identical. For the 15~M$_\odot$
star, where dynamic beta-equilibrium is achieved in the simulation
(see lower right panel of Fig.~\ref{fig:presn}), all the $Y_e$ values
are almost identical for times around $10^4$~s. For the 25~M$_\odot$
star, dynamic beta-equilibrium is never reached and consequently our
equilibrium $Y_e$ is slightly smaller.

Our equilibrium electron fraction represents a static equilibrium and
is obtained independently of the weak-interaction rates. Such an
equilibrium is reached if the temperature and density are kept constant
long enough. There are two points that a set of
weak-interaction rates has to fulfil to be used in presupernova
models: 1) if dynamic beta-equilibrium is reached, the electron
fraction should be very similar to our equilibrium electron fraction;
2) the electron fraction in the presupernova models can not drop below
the equilibrium value we calculate. The older presupernova models of
\cite{Woosley.Weaver:1995} do not fulfil these two
constraints. \cite{Aufderheide.Fushiki.ea:1994b,
 Heger.Woosley.ea:2001} suggested that older models needed to be
recomputed with consistent set of weak-interaction rates. The new
results \cite[]{Heger.Woosley.ea:2001} based on better determination
of the rates satisfy the two constraints. This test should also apply
to future compilations of weak-interaction rates and their
implementation in presupernova simulations.

\subsection{Proto-neutron star}
\label{sec:proto-ns}
Iron core collapse marks the end of the presupernova phase. When the
collapse starts the dynamical evolution is very fast compared to the
weak interaction timescale, thus the assumption of equilibrium is not
valid. But when collapse continues, neutrinos are trapped due to the
increasingly higher central densities, and thermal weak equilibrium
can be reached again. However, neutrinos are trapped and their
chemical potentials are not negligible anymore ($\mu_{\nu} \neq
0$). Therefore, our results give just a rough estimation of the
electron fraction during collapse, but nevertheless are useful to have
an idea of the composition in the collapsing material. In
Fig.~\ref{fig:ye} the trajectory of the collapse is in a region where
almost only heavy neutron-rich nuclei and neutrons are present
(Fig.~\ref{fig:nuclei}). The amount of alpha particles
(Fig.~\ref{fig:xa}) and light nuclei (Fig.~\ref{fig:lights}) is
negligible.

\begin{figure}
 \includegraphics[width=0.9\linewidth]{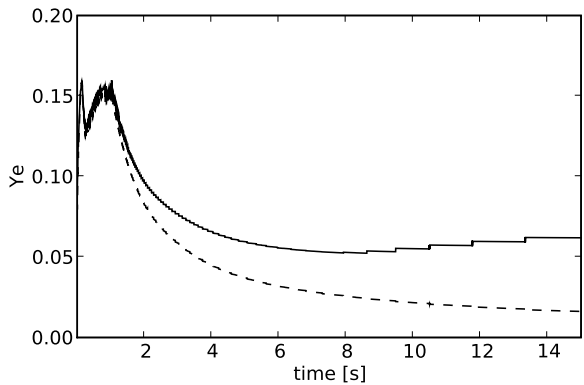}
 \caption{The evolution of the electron fraction at the proto-neutron
   star surface is shown. The solid line is obtained by assuming NSE
   and thermal weak equilibrium, while the dashed line results from
   the hydrodynamical simulation of the supernova explosion and the
   subsequent neutrino-driven wind of
   \cite{arcones.janka.scheck:2007}.}
 \label{fig:protons}
\end{figure}

When central density gets above the density of nuclear matter,
collapse stops and falling matter bounces back leading to the
formation of a shock. The way this shock propagates through the outer
layers of the star in a supernova explosion is still a problem under
discussion \cite[]{Janka.Langanke.ea:2007}.  But it is known that a
compact object forms in the center and cools by neutrino
emission. Neutrinos start to decouple from matter at the
neutrinosphere that is placed in the outer layers of the newly born
proto-neutron star. In this region we assume still that $\mu_{\nu}
\approx 0$. Also NSE is satisfied because the temperatures are high
enough, $T \approx 50 \cdot 10^9$~K. We compare our equilibrium
electron fraction with the one obtained in hydrodynamical simulations
of core-collapse supernovae \cite[]{arcones.janka.scheck:2007}.
Figure~\ref{fig:protons} shows the post-bounce evolution of the
electron fraction. The dashed line is the result of the hydrodynamical
simulation of \cite{arcones.janka.scheck:2007} for their standard case
(M15l1r1), and solid line is the equilibrium electron fraction
computed for same density and temperature evolution as for the dashed
line. There is a clear discrepancy between the two electron fractions
that increases with time. This disagreement comes from the difference
in the composition \cite[see Fig.~4
in][]{arcones.matinezpinedo.etal:2008}.  The equations of state that
are generally used in supernova simulations include only nucleons,
alpha particles, and a representative heavy nucleus. In our
calculation of NSE with beta equilibrium, we take into account all
nuclei. Figure~\ref{fig:lights} shows that the abundances of light
elements ($A<4$) are not negligible in the outer layers of the
proto-neutron star. Similar results are found with other EoS, for
example \cite{horowitz.schwenk:2006} and
\cite{Typel.Roepke.etal:2009}.  Light nuclei are formed at expenses of
the few free protons. This produces also a reduction of the
antineutrino absorption and electron captures, which leads to an
increase of the electron fraction compared to the case where only
neutrons and protons are present (see Eq.~(\ref{eq:dotye})).

These light nuclei have an impact on the properties of neutrinos
emitted during the first seconds after bounce
(\cite{arcones.matinezpinedo.etal:2008}).
\cite{sumiyoshi.roepke:2008} have also pointed out the influence of
the light elements during the post-bounce evolution. These
calculations were both done in a post-processing step, therefore the
real effect of light elements in the dynamics of the supernova
evolution is still an open issue that has to be studied in detail by
using an equation of state that includes light nuclei in supernova
simulations.

\subsection{Type Ia supernovae and Accretion-induced collapse}
\label{sec:typeIa}

Other astrophysical scenarios where the electron fraction plays an
important role involve the evolution of white dwarfs in binary
systems. A white dwarf accreting up to the Chandrasekhar limit can
have two possible outcomes. If the density and temperature become
sufficiently high to ignite explosive nuclear burning, then the white
dwarf may become a type Ia supernova. But it could also happen that
electron capture in the ash reduces central temperature and pressure,
and the white dwarf instead collapses to a neutron star. This latter
outcome is known as ``accretion-induced collapse'', or AIC
\cite[]{Canal.Isern.ea:1990, Woosley.Baron:1992, Fryer.Benz.ea:1999}.
Which outcome occurs depends on the ignition density which is in turn
determined by the initial white dwarf properties (mass, composition,
and temperature) and on the accretion rate
\cite[]{Canal.Isern.ea:1990}.

The outcome also depends on the weak interaction rates employed in the
study and the multi-dimensional character of the explosion. Burning
raises the temperature at constant pressure and, if $Y_e$ is constant,
that must reduce the density. The density inversion behind the flame
drives a Rayleigh-Taylor instability that greatly enhances the advance
of the burning front. However, since the pressure is mostly due to
relativistic electrons, a decrease in $Y_e$ due to electron capture in
the ash means that the post-burning density will be larger (since
pressure depends on the product $\rho Y_e$). If $Y_e$ decreases
sufficiently, the density beneath the flame may even rise, thus
suppressing the Rayleigh-Taylor instability \cite[see Sect.~4
in][]{Timmes.Woosley:1992}. If $Y_e$ decreases in a sufficient
fraction of the white dwarf before the (now slowly moving) flame makes
its way out to regions where the density is lower and electron capture
is reduced, the white dwarf collapses. At densities approaching
10$^{10}$ g cm$^{-3}$ electron capture is so efficient that it would
drive $Y_e$ to very low values were it not for the inhibiting effect
of beta-decay. Yet there are, so far, no accurate calculations of
beta-decay rates for $Y_e$ below about 0.41. Having a reliable {\sl
  lower bound} to $Y_e$ at a given density will thus be useful in
future simulations.

\begin{figure}
   \includegraphics[width=0.9\linewidth]{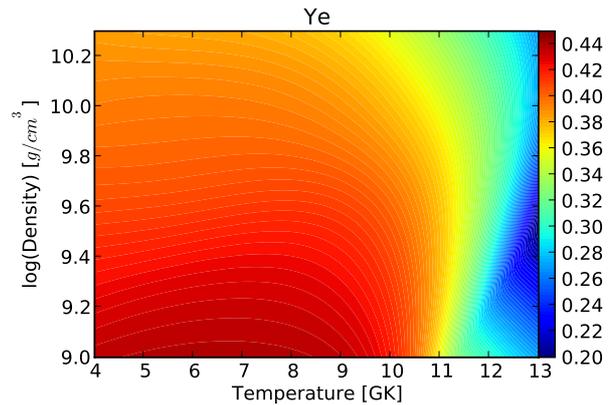}\\
   \caption{The dynamic beta-equilibrium electron fraction
     (Eq.~(\ref{eq:dotyedbe})) is shown for the range of temperatures and
     densities relevant in type Ia supernovae and AICs. In
     Table~\ref{tab:ye} numerical values of the equilibrium electron
     fraction are given.}
 \label{fig:ye_small}
\end{figure}

During this phase of the white dwarf evolution, temperatures are high
and densities relatively low. Consequently, the produced neutrinos
escape. Under such conditions, the lower limit of the electron
fraction is thus obtained from dynamic beta-equilibrium,
Eq.~(\ref{eq:dotyedbe}). If thermal weak equilibrium were assumed,
electron captures would compete with antineutrino absorptions instead
of $\beta^-$ decay, as expected when neutrinos escape.
Figure~\ref{fig:ye_small} shows the dynamic beta-equilibrium electron
fraction for the range of temperatures and densities relevant for AIC
and type Ia supernovae. In Tab.~\ref{tab:ye} some numerical values of
the $Y_e$ in the figure are given (a complete table is available on
request).

The behaviour of the electron fraction is very different at high and
low temperatures (Fig.~\ref{fig:ye_small}). At low temperatures the
electron fraction decreases continuously with density. While a high
temperatures ($T\gtrsim 12$~GK) first decreases and then increases.
This seems unintuitive but it can be understood considering the change
in composition. At low densities the composition is neutrons, protons,
and alpha-particles. With increasing density heavy nuclei are present
and their $\beta^-$ decay produces an increase of $Y_e$. At even
higher densities, electron capture dominates and $Y_e$ decreases
again.

\begin{table*}
   \begin{tabular}{r|cccccccccccc}
     \hline \hline
     $\rho_9=$ &    1.02 &  2.11 &  3.03 & 4.05 &  5.04  &    7.24 &  9.00 &  14.96  &  20.00 \\
    \hline 
     $T_9=$  13.03 &   0.269 &  0.224  &  0.202 & 0.207 & 0.229  & 0.256  & 0.265  &0.270  & 0.266 \\
     $T_9=$  12.02 &   0.299 &  0.282  &  0.307 & 0.319 & 0.324  & 0.326  & 0.325  &0.314  & 0.305 \\
     $T_9=$  11.09 &   0.351 &  0.371  &  0.373 & 0.371 & 0.369  & 0.362  & 0.357  &0.342  & 0.332 \\
     $T_9=$  10.03 &   0.419 &  0.410  &  0.405 & 0.399 & 0.394  & 0.385  & 0.379  &0.365  & 0.355 \\
     $T_9=$  9.075 &   0.434 &  0.423  &  0.416 & 0.409 & 0.404  & 0.395  & 0.390  &0.377  & 0.369 \\
     $T_9=$  8.044 &   0.437 &  0.428  &  0.420 & 0.413 & 0.407  & 0.399  & 0.395  &0.385  & 0.377 \\
     $T_9=$  7.058 &   0.438 &  0.428  &  0.420 & 0.412 & 0.406  & 0.400  & 0.396  &0.388  & 0.381 \\
     $T_9=$  6.010 &   0.438 &  0.427  &  0.418 & 0.410 & 0.405  & 0.399  & 0.397  &0.389  & 0.383 \\
     \hline \hline
   \end{tabular}
   \caption{Dynamic beta-equilibrium electron fraction for  density values given in the first row and temperature values in the first column. The temperature is in units of $10^9$~K and the density in $10^9$~g~cm$^{-3}$. These temperature and density intervals correspond to the Fig.~\ref{fig:ye_small}.}
   \label{tab:ye} 
\end{table*}

Finally, we notice that the electron fraction and thus
weak-interaction rates are important for determining the composition
of the ejecta in typical type Ia supernova. In these events the
electron captures are very important during the flame propagation
because they lead to a reduction of the electron fraction which is the
parameter that controls the isotopic composition of the ejecta.

\section{Conclusions}
\label{sec:conclusions}
Weak-interaction rates determine the evolution of the electron
fraction, which is a key parameter affecting the composition and
dynamics of stars in the late stages of stellar evolution and
supernovae.  Here we have presented an interesting limiting case that
can be used to test the implementation of theoretical weak-interaction
rates in stellar evolution simulations and to place lower bounds on
the $Y_e$ in the explosion of white dwarfs. Our results depend only on
chemical potentials and therefore are independent of the
weak-interaction rates.

These equilibrium electron fractions can be compared to those found in
presupernova models, where weak-interaction rates play an important
role \cite[]{Heger.Langanke.ea:2001}. The presupernova phase precedes
the collapse of massive stars, therefore temperatures get high enough
that NSE is reached, but the dynamical evolution is too fast in most
cases for beta equilibrium to be achieved. We find a lower limit for
electron fraction and show that the $Y_e$ in the old presupernova
models of \cite{Woosley.Weaver:1995} drops below it. In these
presupernova models beta decay rates of \cite{Hansen:1968,
  Mazurek.Truran.Cameron:1974} were used. These rates are too low
compared to FFN and LMP. Similar models calculated using more recent
weak rates \cite[]{Heger.Woosley.ea:2001} lead to an electron fraction
that stays always above or equal to our lower limit. Moreover, for the
15~$M_{\odot}$ model where the rates predict dynamic beta-equilibrium,
we find that their electron fraction agrees with the equilibrium value
of our calculations.

The accretion onto white dwarfs approaching the Chandrasekhar mass
could lead to type Ia supernovae or to AIC. The final outcome depends
on the competition between explosive ignition and electron captures
\cite[]{Canal.Isern.ea:1990}. Therefore, the electron fraction is a
key parameter. In this environment densities and temperatures drive
the NSE composition towards neutron-rich nuclei whose weak-interaction
rates are not yet known from theoretical models. Here we give a
reliable lower limit for $Y_e$, that can be used in simulations (data
are available on request).

In addition, our calculations show that the amount of light elements
($A<4$) in the outer layer of the proto-neutron star is not
negligible. After supernova explosion a hot proto-neutron star is
born and cools emitting neutrinos.  The surface of the proto-neutron
star, where neutrinos decouple from matter, is hot enough to assume
NSE and its evolution is rather slow, thus beta equilibrium is also
satisfied. Changes in the composition of this region have an impact on
neutrino properties that could affect the nucleosynthesis
\cite[]{arcones.matinezpinedo.etal:2008} and the explosion
\cite[]{sumiyoshi.roepke:2008}.

\begin{acknowledgements}
  We thank A.~ Marek and F.~R\"opke for providing the collapse and
  type Ia supernova trajectories, and H.~-Th.~Janka, K.~Langanke,
  I.~Seitenzahl for valuable discussions. The work of A.~Arcones and
  G.~Mart\'inez-Pinedo was supported by the Deutsche
  Forschungsgemeinschaft through contract SFB 634 and by the ExtreMe
  Matter Institute EMMI. S.~E.~Woosley was supported by the US NSF
  (AST-0909129), the University of California Office of the President
  (09-IR-07-117968-WOOS), and the DOE SciDAC Program
  (DE-FC-02-06ER41438).  L.~Roberts was supported by an NNSA/DOE
  Stewardship Science Graduate Fellowship (DE-FC52-08NA28752) and the
  University of California Office of the President
  (09-IR-07-117968-WOOS).
\end{acknowledgements}

%\bibliographystyle{aa}
%\bibliography{bibliography}

\end{document}